\documentclass[12pt]{iopart}
\usepackage{graphicx}

\begin{document}

\title[Gravitational wave detection using high precision pulsar observations]{Gravitational wave detection using high precision pulsar observations}

\author{G. Hobbs}

\address{Australia Telescope National Facility, CSIRO, P.O. Box 76, Epping, NSW 1710 Australia.}
\ead{george.hobbs@csiro.au}
\begin{abstract}
Pulsar timing experiments are reaching sufficient sensitivity to detect a postulated stochastic gravitational wave background generated by merging supermassive black hole systems in the cores of galaxies.  We describe the techniques behind the pulsar timing detection method, provide current upper bounds on the amplitude of any gravitational wave background, describe theoretical models predicting the existence of such a background and highlight new techniques for providing a statistically rigorous detection of the background.
\end{abstract}


\section{Introduction}

It is now possible to make timing observations of millisecond pulsars to a precision of $\sim 100$\,ns.  Such exquisite timing precision allows the pulsar astrometric, spin and orbital parameters to be determined with great accuracy.  This has recently allowed us to limit the rate of change of the gravitational constant, $|\dot{G}/G| < 2.1\times10^{-12}$\,yr$^{-1}$, and to obtain the most accurate astrophysical distance estimate outside of the Solar System (Verbiest et al. 2007, submitted to ApJ).  Multi-frequency pulsar observations have allowed us to study the Solar corona (You et al. 2007, in press) and density fluctuations in the interstellar plasma (You et al. 2007b)\nocite{yhc+07}. One of the most exciting applications of such data-sets is to search for the signatures of gravitational waves (GWs) passing over the Earth.    This is the main goal of the Parkes Pulsar Timing Array (PPTA) project\footnote{http://www.atnf.csiro.au/research/pulsar/ppta}, which aims to observe 20 millisecond pulsars with a timing precision close to 100\,ns over more than five years.

\begin{figure}
\begin{center}\includegraphics[width=8cm,angle=-90]{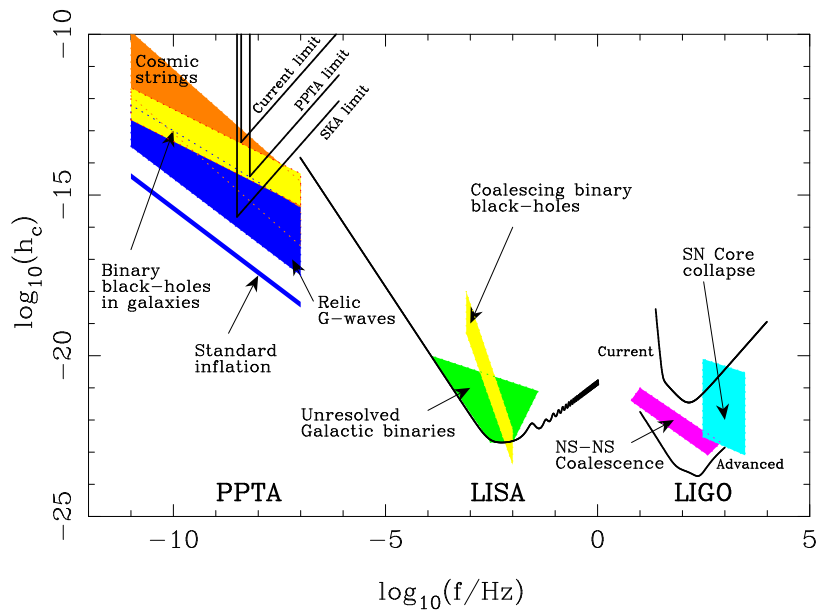}\end{center}
\caption{Characteristic strain sensitivity for existing and proposed GW detectors as a function of GW frequency. Predicted signal levels from various astrophysical sources are shown.}\label{fg:sens}
\end{figure}

The existence of a GW signal can be identified by analysis of pulsar timing residuals.  In brief, the pulsar timing model (see, e.g. Edwards, Hobbs \& Manchester 2006)\nocite{ehm06} allows the observed arrival times of pulses from a radio pulsar to be compared with a model of the pulsar's astrometric, orbital and rotational parameters.  The differences between the actual arrival times and the predicted arrival times are the ``pulsar timing residuals".  Any features observed in the timing residuals indicate the presence of unmodelled effects which may include calibration errors, (additional) orbital companions, spin-down irregularities, or GWs. The induced pulsar timing residuals due to a GW signal were first calculated by Sazhin (1978)\nocite{saz78} and Detweiler (1979)\nocite{det79}.  They showed that a GW signal causes a fluctuation in the observed pulse frequency $\delta \nu/\nu$ which affects pulsar timing residuals at time $t$ from the initial observation as
\begin{equation}
R(t) = -\int_0^t\frac{\delta \nu(t)}{\nu}dt.
\end{equation}
For a given pulsar and GW source, this effect is only dependent upon the characteristic strain at the pulsar and at the Earth. The GW strains evaluated at the positions of multiple pulsars will be uncorrelated, whereas the component at the Earth will lead to a correlated signal in the timing residuals of all pulsars. As pulsars are typically observed every few weeks for many years and standard pulsar timing techniques absorb any low-frequency GWs by fitting for the pulsar's spin-down\footnote{As the intrinsic pulsar spin-period and rate of spin-down is unknown, a quadratic curve is always fitted and subsequently removed from the pulsar timing residuals.}, pulsar timing experiments are sensitive to GW signals in the ultra low frequency ($f \sim 10^{-9}$\,Hz) band. This technique is therefore complementary to other GW detection methods such as the Laser Interferometer Space Antenna (LISA) and ground based interferometer systems (such as LIGO and VIRGO), which are sensitive to higher frequency GWs ($f \sim 10^{-2.5}$ and $f\sim 10^{2.5}$\,Hz respectively). In Figure~\ref{fg:sens} we plot the sensitivity of the pulsar timing experiments, LISA and LIGO and the expected GW sources.  For the pulsar timing experiments, maximum sensitivity is reached at $f = 1/T$ where $T$ is the data span. As described above, the pulsar experiments are not sensitive to GWs with $f < 1/T$ and, for white pulsar timing residuals, the upper bound on the characteristic strain increases as $f^{3/2}$ for $f > 1/T$.

In this paper we review the potential sources of GWs (\S\ref{sec:sources}) before describing how upper bounds on the amplitude of any existing background can be calculated and highlighting the astrophysical implications of these bounds. In \S\ref{sec:detect} we describe how a stochastic GW background may be detected.  Finally, we highlight some practical issues that need to be addressed in achieving our sensitivity goal (\S\ref{sec:prac}) and describe possible future experiments (\S\ref{sec:future}).

\section{Potential sources of gravitational waves}\label{sec:sources}

Theoretical models predict that pulsar timing experiments will be sensitivity to burst GW sources, individual sources and to a stochastic background.

\subsection{Single sources}

\begin{figure}
\begin{center}\includegraphics[width=5cm,angle=-90]{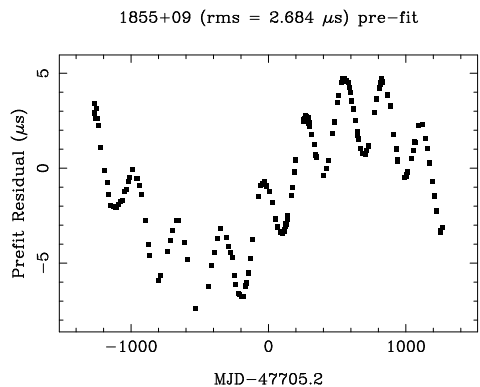}\end{center}
\caption{Simulation of the induced timing residuals for PSR~B1855+09 due to the postulated binary black hole system in the radio galaxy 3C66B.}\label{fg:simul}
\end{figure}

Recently Sudou et al. (2003)\nocite{simt03} reported the possible discovery of a binary supermassive black hole system in the radio galaxy 3C66B.  As shown by Jenet et al. (2004)\nocite{jllw04}, the GW radiation from this postulated system would have produced clearly detectable fluctuations in the timing residuals of existing pulsar data-sets.  The non-detection of the expected GW signal (see Figure~\ref{fg:simul}) ruled out the postulated system with 95\% confidence. However, detecting (or limiting the existence of) individual supermassive binary black hole systems is necessary for testing theories of hierarchical galaxy formation. Order-of-magnitude calculations demonstrate that a system with a chirp mass of $10^{9}$M$_\odot$ and an orbital period of 10\,yr, at a distance of 20\,Mpc will induce a sinusoidal signal in the timing residuals with amplitude $\sim 100$\,ns at a frequency of twice the orbital frequency.

\subsection{Burst sources}

Sources of burst GW emission that may be detectable with current and proposed pulsar timing experiments include 1) the formation of supermassive black holes (Thorne \& Braginskii 1976)\nocite{tb76}, which lead to a day-long burst of radiation, 2) highly eccentric supermassive black hole binaries (Enoki \& Nagashima 2007)\nocite{en07}, 3) close encounters of massive objects (Kocsis et al. 2006)\nocite{kgm06} and 4) cosmic string cusps (Damour \& Vilenkin 2001)\nocite{dv01}.  The sensitivity of a pulsar timing array experiment to burst GW emission depends upon the number of pulsars in the array, the timing precision, number of observations, the pulsar positions on the sky and the coordinates of the GW burst source.  With our final data-sets we expect to achieve an angular resolution of approximately 30$^\circ$ (Lommen, private communication). By sharing data with timing array projects being carried out in the Northern hemisphere we expect to improve both the angular resolution of our detector and our sensitivity to burst sources in the Northern Hemisphere.

\subsection{Stochastic background}

\begin{figure}
\begin{center}\includegraphics[width=5cm,angle=-90]{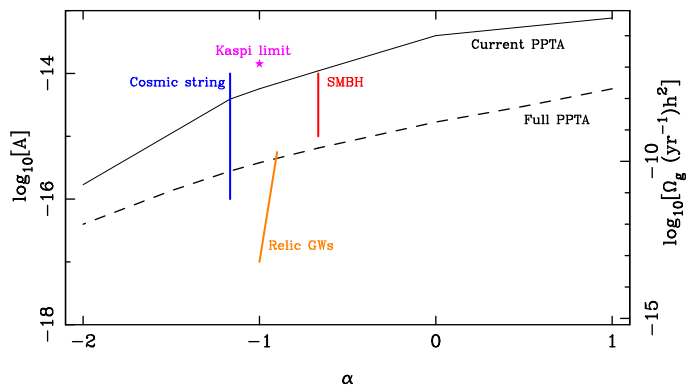}\end{center}
\caption{The solid line indicates our current limits on GW backgrounds of specified spectral exponent.  The dashed line underneath indicates the expected limits within 5 years.  The near vertical lines give predicted background amplitudes.  The ``star"-symbol indicates the earlier limit obtained by Kaspi et al. (1994).}\label{fg:a_upper}
\end{figure}

A stochastic background of GWs due to binary supermassive black holes (Jaffe \& Backer 2003\nocite{jb03}, Wyithe \& Loeb 2003\nocite{wl03}, Enoki et al. 2004\nocite{eins04}, Sesana et al. 2004\nocite{shmv04}), cosmic strings or relic GWs from the big bang (Maggiore 2000\nocite{mag00}) are all potentially detectable.  In most models, the GW strain spectrum,  $h_c(f)$, can be represented by a power-law in the GW frequency, $f$,
\begin{equation}
h_c(f) = A\left(\frac{f}{\rm yr^{-1}}\right)^\alpha
\end{equation}
where the spectral exponent $\alpha = -2/3$, $-1$ and $-7/6$ for likely GW backgrounds due to coalescing black hole binaries, cosmic strings and relic GWs respectively (Figure~\ref{fg:a_upper}).  The energy density of the background per unit logarithmic frequency interval can be written as
\begin{equation}
\Omega_{\rm GW}(f) = \frac{2}{3}\frac{\pi^2}{H_0^2}f^2h_c(f)^2
\end{equation}
where $H_0$ is the Hubble constant.

Jenet et al. (2006)\nocite{jhs+06} introduced a method to place an upper bound on $A$ for a given $\alpha$ using actual pulsar timing residuals. In contrast with earlier techniques (e.g.  Kaspi et al. 1994\nocite{ktr+94}) this method used the \textsc{tempo2} software package (Hobbs, Edwards \& Manchester 2006)\nocite{hem06} to account for all the fitting procedures undertaken during the pulsar timing process.  Using observations from the PPTA project along with archival data from the Arecibo telescope, Jenet et al. (2006)\nocite{jhs+06} obtained an upper limit on the energy density per unit logarithmic frequency interval of $\Omega_{\rm GW}[1/(8{\rm yr})]h^2 \leq 2\times10^{-8}$ for $\alpha=-2/3$, which corresponds to $A \leq10^{-14}$ (see Figure~\ref{fg:a_upper} and the ``Current limit'' in Figure~\ref{fg:sens}).  This is the most stringent limit to date on the existence of a GW background in the nano-Hertz frequency band and has astrophysical and cosmological implications.  For instance, this limit was used to 1) constrain the merger rate of massive black hole systems at high redshift, 2) rule out some relationships between the black hole mass and galactic halo mass, 3) constrain the rate of expansion in the inflationary era and 4) provide an upper bound on the dimensionless tension of cosmic strings.

We have recently shown that the Jenet et al. (2006)\nocite{jhs+06} technique described above suffers from two limitations.  First, it does not give the lowest possible upper bound and secondly, it relies on the observed spectrum being white.  Unfortunately, even for millisecond pulsars, it is unusual for the timing residuals not to be affected by unexplained phenomena, such as pulsar ``timing noise" (see e.g. Hobbs, Lyne \& Kramer 2006)\nocite{hlk06}.  A new method has now been developed and will be described in a subsequent paper.  Our new method is applicable to any pulsar timing data-set and will lead to significantly improved bounds on the GW background amplitude.

\section{Detecting a background}\label{sec:detect}

\begin{figure}
\begin{center}\includegraphics[width=7cm,angle=-90]{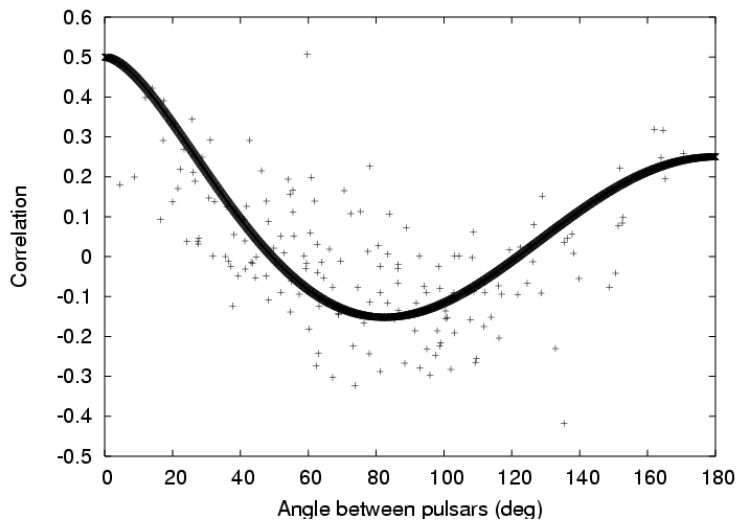}\end{center}
\caption{The expected correlation in the timing residuals of pairs of pulsars as a function of angular separation for an isotropic GW background.}\label{fg:corr}
\end{figure}

Jenet et al. (2005)\nocite{jhlm05} developed a technique for making a definitive detection of an isotropic, stochastic, background of GWs by searching for correlated signals between pulsar data-sets.  For each pair of pulsars, the zero-lag correlation between the respective timing residuals can be calculated.  The expected correlation as a function of angle between the pulsars for timing residuals dominated by a GW background was first calculated by Hellings \& Downs (1983)\nocite{hd83} and is shown in Figure~\ref{fg:corr}\footnote{This correlation curve assumes the general theory of relativity. We are calculating the equivalent curves for other theories of gravity.  These curves and their implications will be discussed elsewhere.}.  Jenet et al. (2005)\nocite{jhlm05} showed, for expected amplitudes of a GW background, that this signal could be unambiguously detected if 20 or more pulsars were observed over a period of 5 years each with an rms timing residual of ~100---500\,ns.

\begin{figure}
\begin{center}\includegraphics[width=10cm]{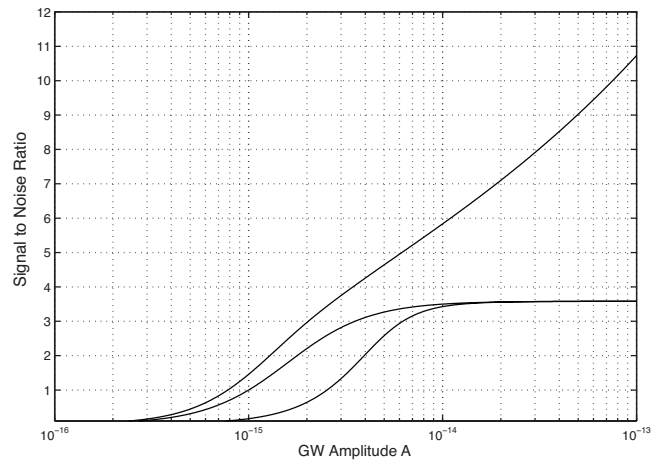}\end{center}
\caption{The sensitivity of the completed PPTA project to a GW background of specified amplitude. The three curves represent no pre-filtering of the data (lowest), using an optimal low-pass filtering technique (middle) and optimally pre-whitening the data-sets (highest).}\label{fg:sig_a}
\end{figure}

We have recently generalised the initial work of Jenet et al. (2005) to produce routines that can be applied to the observed pulsar timing residuals (with the actual sampling, data lengths etc.).  This new technique also increases the significance of any possible detection, by optimally pre-whitening the data-sets. An example of the improvement possible by pre-whitening the timing residuals is shown in Figure~\ref{fg:sig_a} where the significance of detection is plotted versus the GW background amplitude for the PPTA goal of timing 20 pulsars with 100\,ns timing precision over a period of five years. Without any pre-filtering of the data we will only achieve a maximum detection significance of $S \sim 3.5$ (corresponding to the sensitivity labelled as the ``PPTA limit'' in Figure~\ref{fg:sens}).  Using optimal whitening schemes this sensitivity can be significantly enhanced. Initial results suggest that if we can reduce our current timing precision for all pulsars by a factor of $\sim 2$ we will be sensitive, within five years, to the maximum amplitude predicted by models of an isotropic stochastic GW background caused by coalescing black-holes in the centres of galaxies\footnote{Note that this limit is similar to the current limits described in \S2.3 that were obtained using a few long data-sets.  We emphasise that, even though limits can be placed using a few data-sets, detecting a GW background requires observations of $\sim$20\,pulsars.}. 

\section{Practical issues}\label{sec:prac}

Pulsar timing is affected by the stability of terrestrial clocks, ephemeris errors and the pulsars themselves.  The goal of the PPTA project, described above,  to obtain data-sets for 20 pulsars with approximately two-weekly sampling with rms timing residuals of $\sigma \sim 100$\,ns over five years, is challenging.  To date we have a few pulsars where we can obtain such precision. For instance, van Straten et al. (2001)\nocite{vbb+01} obtained $\sigma = 130$\,ns over 40 months of observing PSR~J0437$-$4715. For the majority of our pulsars $\sigma \sim 1\mu s$.  We have many possibilities for decreasing these timing residuals, including:

\begin{itemize}
\item{\emph{correcting the data-sets for dispersion measure variations}: You et al. (2007) showed that for PSR~J1939+2134, $\sigma$ decreased from 290\,ns to 190\,ns after correction.}
\item{\emph{improving instrumentation}: We have recently developed new digital filterbank systems for the Parkes telescope.  The recently commissioned system provides 1\,GHz of bandwidth at an observing frequency of 3\,GHz.  We are also developing a new wide-bandwidth coherent dedispersion system.}
\item{\emph{new processing techniques}:  We are exploring new methods to calibrate our data-sets which should lead to at least a factor of 2 improvement in our rms residuals.  We have recently produced a new pulsar timing package, \textsc{tempo2} (Hobbs, Edwards \& Manchester 2006)\nocite{hem06}, that  is accurate (for known physics) at the 1\,ns level.}
\end{itemize}

\section{Future possibilities}\label{sec:future}

The sensitivity of a pulsar timing array to a stochastic background is proportional to the average timing precision, the square root of the number of observations and to the number of pulsars in the array. 
This paper has concentrated on the Parkes pulsar timing array project.  However, significant improvements will be made to the timing array sensitivity to GW sources by combining our data-sets with Northern Hemisphere pulsar observations.  The North American pulsar timing array project (NanoGrav) has been observing a sample of pulsars using the Arecibo and GreenBank observatories.  These observations provide more than 20\,years of timing PSRs B1855+09 and B1937+21 and were used to provide the first stringent limits on the stochastic gravitational wave background (e.g. Kaspi et al. 1994) and to rule out the postulated binary black hole system in the radio galaxy 3C66B (Jenet et al. 2004).  The European Pulsar Timing Array (EPTA) team have recently started to obtain data on a large sample of pulsars using the four major European telescopes at Jodrell Bank, Effelsberg, Nancay and Westerbork.  We are currently developing techniques to combine observations from different observatories and will soon produce new limits on the existence of a GW background using all available data.    

Figure~\ref{fg:sens} shows the sensitivity that should be achievable using the Square Kilometer Array (SKA) telescope hoped to be built by the year 2020.  This sensitivity curve assumes an SKA timing array project that could observe 100 millisecond pulsars with rms timing residuals around 50\,ns for a period of 10\,yr. Provided intrinsic pulsar timing noise does not dominate the effect of the stochastic GW background then this limit corresponds to a detection limit at 3nHz of $\Omega_{gw} \sim 10^{-13}$.  Pre-whitening procedures have the potential to greatly decrease this limit.

\section{Conclusion}

It is possible that gravitational waves will be detected within the next decade by world-wide pulsar timing array projects.  As a by-product of these investigations stringent checks will also be placed on terrestrial time standards and the solar system ephemeris.  Regular dual-frequency observations of multiple pulsars will also provide valuable information about the interstellar medium.  Using a pulsar array as a gravitational wave detector is complimentary to other searches that are attempting to detect much higher-frequency gravitational waves.

\subsection{Acknowledgments}

This work is undertaken as part of the Parkes Pulsar Timing Array project, which is a collaboration between the ATNF, Swinburne University and the University of Texas, Brownsville.  The author wishes to than R. Manchester, W. Coles and F. Jenet for helping to improve this paper. The Parkes radio telescope is part of the Australia Telescope, which is funded by the Commonwealth of Australia for operation as a National Facility managed by CSIRO.

\section*{References}

\bibliographystyle{aipproc} 
\bibliography{journals,modrefs,psrrefs,crossrefs}

\begin{thebibliography}{20}
\expandafter\ifx\csname natexlab\endcsname\relax\def\natexlab#1{#1}\fi
\providecommand{\enquote}[1]{``#1''}
\expandafter\ifx\csname url\endcsname\relax
  \def\url#1{\texttt{#1}}\fi
\expandafter\ifx\csname urlprefix\endcsname\relax\def\urlprefix{URL }\fi
\providecommand{\eprint}[2][]{\url{#2}}

\bibitem{yhc+07}
X.~P. {You}, G.~{Hobbs}, W.~A. {Coles}, R.~N. {Manchester}, R.~{Edwards},
  M.~{Bailes}, J.~{Sarkissian}, J.~P.~W. {Verbiest}, W.~{van Straten},
  A.~{Hotan}, S.~{Ord}, F.~{Jenet}, N.~D.~R. {Bhat}, and A.~{Teoh},
  \emph{MNRAS} \textbf{378}, 493 (2007).

\bibitem{ehm06}
R.~T. {Edwards}, G.~B. {Hobbs}, and R.~N. {Manchester}, \emph{MNRAS}
  \textbf{372}, 1549--1574 (2006).

\bibitem{saz78}
M.~V. Sazhin, \emph{Sov. Astron.} \textbf{22}, 36 (1978).

\bibitem{det79}
S.~Detweiler, \emph{ApJ} \textbf{234}, 1100 (1979).

\bibitem{simt03}
H.~{Sudou}, S.~{Iguchi}, Y.~{Murata}, and Y.~{Taniguchi}, \emph{Science}
  \textbf{300}, 1263--1265 (2003).

\bibitem{jllw04}
F.~A. Jenet, A.~Lommen, S.~L. Larson, and L.~Wen, \emph{ApJ} \textbf{606},
  799--803 (2004).

\bibitem{tb76}
K.~S. {Thorne}, and V.~B. {Braginskii}, \emph{ApJ} \textbf{204}, L1--L6 (1976).

\bibitem{en07}
M.~{Enoki}, and M.~{Nagashima}, \emph{Progress of Theoretical Physics}
  \textbf{117}, 241--256 (2007).

\bibitem{kgm06}
B.~{Kocsis}, M.~E. {G{\'a}sp{\'a}r}, and S.~{M{\'a}rka}, \emph{ApJ}
  \textbf{648}, 411--429 (2006).

\bibitem{dv01}
T.~Damour, and A.~Vilenkin, \emph{Phys. Rev. D} \textbf{71}, 063510 (2001).

\bibitem{jb03}
A.~H. Jaffe, and D.~C. Backer, \emph{ApJ} \textbf{583}, 616--631 (2003).

\bibitem{wl03}
J.~S.~B. {Wyithe}, and A.~ {Loeb}, \emph{ApJ} \textbf{590}, 691-706 (2003).

\bibitem{eins04}
M.~{Enoki}, K.~T. {Inoue}, M.~{Nagashima}, and N.~{Sugiyama}, \emph{ApJ}
  \textbf{615}, 19--28 (2004).

\bibitem{shmv04}
A.~{Sesana}, F.~{Haardt}, P.~{Madau}, and M.~{Volonteri}, \emph{ApJ}
  \textbf{611}, 623--632 (2004).

\bibitem{mag00}
M.~Maggiore, \emph{Phys. Rep.} \textbf{331}, 283--367 (2000).

\bibitem{jhs+06}
F.~A. {Jenet}, G.~B. {Hobbs}, W.~{van Straten}, R.~N. {Manchester},
  M.~{Bailes}, J.~P.~W. {Verbiest}, R.~T. {Edwards}, A.~W. {Hotan}, J.~M.
  {Sarkissian}, and S.~M. {Ord}, \emph{ApJ} \textbf{653}, 1571--1576 (2006).

\bibitem{ktr+94}
V.~M. {Kaspi}, J.~H. {Taylor} and M.~{Ryba}, \emph{ApJ} \textbf{428}, 713-728 (1994)

\bibitem{hem06}
G.~B. {Hobbs}, R.~T. {Edwards}, and R.~N. {Manchester}, \emph{MNRAS}
  \textbf{369}, 655--672 (2006{\natexlab{a}}).

\bibitem{hlk06}
G.~{Hobbs}, A.~{Lyne}, and M.~{Kramer}, \emph{Chin. J. Astron. Astrophys.,
  Suppl. 2} \textbf{6}, 169--175 (2006{\natexlab{b}}).

\bibitem{jhlm05}
F.~A. {Jenet}, G.~B. {Hobbs}, K.~J. {Lee}, and R.~N. {Manchester}, \emph{ApJ}
  \textbf{625}, L123--L126 (2005).

\bibitem{hd83}
R.~W. Hellings, and G.~S. Downs, \emph{ApJ} \textbf{265}, L39 (1983).

\bibitem{vbb+01}
W.~van Straten, M.~Bailes, M.~Britton, S.~R. Kulkarni, S.~B. Anderson, R.~N.
  Manchester, and J.~Sarkissian, \emph{Nature} \textbf{412}, 158--160 (2001).

\end{thebibliography}

\end{document}